\documentclass[12pt,thmsa]{article}
\usepackage{amssymb}

\input tcilatex
\QQQ{Language}{American English}

\begin{document}

\begin{center}
\emph{Resolving the Degeneracy: Experimental tests of the New Self Creation Cosmology}

\emph{and a heterodox prediction for Gravity Probe B}

Garth A Barber

Kingswood Vicarage, 

Woodland Way, Tadworth, Surrey KT20 6NW, England.

Tel: +44 01737 832164\qquad e-mail: garth.barber@virgin.net
\end{center}

\begin{center}
\allowbreak \textit{Abstract}
\end{center}

The new theory of Self Creation Cosmology has been shown to yield a concordant cosmological 
solution that does not require inflation, exotic non-baryonic Dark matter or Dark Energy to fit 
observational constraints. In vacuo there is a conformal equivalence between this theory and 
canonical General Relativity and as a consequence an experimental degeneracy exists as the two 
theories predict identical results in the standard tests. However, there are three 
definitive experiments that are able to resolve this degeneracy and distinguish between the two 
theories. Here these standard tests and definitive experiments are described. One of the 
definitive predictions, that of the geodetic precession of a gyroscope, has just been measured 
on the Gravity Probe B satellite, which is at the present time of writing in the data processing 
stage. This is the first opportunity to falsify Self Creation Cosmology. The theory predicts a 
'frame-dragging' result equal to GR but a geodetic precession of only 2/3 the GR value. When 
applied to the Gravity Probe B satellite, Self Creation Cosmology predicts an E-W 
gravitomagnetic/frame-dragging precession, equal to that of GR, of 40.9 milliarcsec/yr but a 
N-S gyroscope (geodetic + Thomas) precession of just 4.4096 arcsec/yr.
\section{Introduction}

\subsection{The Principles of the Theory}

\subsubsection{Self Creation Cosmology Theories}

This author has described a new Self Creation Cosmology (SCC) (Barber, 2002a) 
with interesting empirical predictions (Barber, 2002b).
This theory superceded two earlier toy theories (SCC1 \& SCC2) (Barber, 1982),
of which SCC1 was discarded as experimentally and internally inconsistent and SCC2 was
subsequently found to be a particular representation of the latest theory. 
All three SCC theories produce continuous creation by modifications of the scalar-tensor 
Brans-Dicke theory (BD), (Brans \& Dicke, 1961) in which the conservation of energy-momentum
is relaxed in order to explore cosmologies in which the matter universe may be created 
out of self-contained gravitational, scalar and matter fields. They have generated
some interest in the literature with approximately 54 citations published over
the last 20 years [see (Barber, 2002b)].

In these theories Mach's Principle (MP) is incorporated by assuming the
inertial masses of fundamental particles are dependent upon their
interaction with a scalar field $\phi $ coupled to the large scale
distribution of matter in motion in a similar fashion as BD. This coupling
is described by a field equation of the simplest general covariant form 
\begin{equation}
\Box \phi =4\pi \lambda T_M^{\;}\text{ ,}  \label{eq2}
\end{equation}
$T_{M\;}^{\;}$ is the trace, ($T_{M\;\sigma }^{\;\;\sigma }$), of the energy
momentum tensor describing all non-gravitational and non-scalar field energy.

\subsubsection{The New Self Creation Cosmology}

In the new theory the BD coupling parameter $\lambda $ was found to be
unity, (Barber, 2002a) and in the spherically symmetric One Body problem 
\begin{equation}
\stackunder{r\rightarrow \infty }{Lim}\phi \left( r\right) =\frac 1{G_N}%
\text{ ,}  \label{eq1a}
\end{equation}
where $G_N$ is the normal gravitational constant measured in Cavendish type
experiments.

In both General Relativity (GR) and BD the equation describing the
interchange of energy between matter and gravitation is,

\begin{equation}
\nabla _\mu T_{M\;\nu }^{\;\mu }=0\text{ ,}  \label{eq1}
\end{equation}
however in all the SCC theories this condition, which arises from the
Equivalence Principle, is relaxed. In the latest SCC theory it was replaced
by the Principle of Mutual Interaction (PMI) in which 
\begin{equation}
\nabla _\mu T_{M\;\nu }^{.\;\mu }=f_\nu \left( \phi \right) \Box \phi =4\pi
f_\nu \left( \phi \right) T_{M\;}^{\;}\text{ ,}  \label{eq7}
\end{equation}
and therefore \textit{in\ vacuo,} \textit{\ } 
\begin{equation}
\nabla _\mu T_{em\quad \nu }^{\quad \mu }=4\pi f_\nu \left( \phi \right)
T_{em}^{\;}=4\pi f_\nu \left( \phi \right) \left( 3p_{em}-\rho _{em}\right)
=0  \label{eq8}
\end{equation}
where $p_{em}$ and $\rho _{em}$ are the pressure and density of an
electromagnetic radiation field with an energy momentum tensor $T_{em\,\mu
\nu }$ and where $p_{em}=\frac 13\rho _{em}$. Thus the scalar field is a
source for the matter-energy field if and only if the matter-energy field is
a source for the scalar field. Although the equivalence principle is
violated for particles, it is not for photons, which still travel through
empty space on (null) geodesic paths.

The effect of the PMI is that particles do not have invariant rest mass. A
second principle, the Local Conservation of Energy, was introduced to
determine the variation in rest mass. It requires a particle's rest mass to
include gravitational potential energy and is described by 
\begin{equation}
m_p(x^\mu )=m_0\exp [\Phi _N\left( x^\mu \right) ]\text{ ,}  \label{eq24}
\end{equation}
where $\Phi _N\left( x^\mu \right) $ is the dimensionless Newtonian
potential and $m_p\left( r\right) \rightarrow m_0$ as $r\rightarrow \infty $.

There is a conformal equivalence between canonical GR and the SCC Jordan Frame that results 
in the geodesic orbits of SCC being identical with GR 
\textit{in vacuo}. Two conformal frames were defined; the Jordan energy
frame [JF(E)], which conserves mass-energy, and the Einstein frame (EF),
which conserves energy momentum. The two conformal frames are related by a
coordinate transformation 
\begin{equation}
g_{\mu \nu }\rightarrow \widetilde{g}_{\mu \nu }=\Omega ^2g_{\mu \nu }\text{
.}  \label{eq9}
\end{equation}
A mass is conformally transformed according to 
\begin{equation}
m\left( x^\mu \right) =\Omega \widetilde{m}_0\text{ ,}  \label{eq33}
\end{equation}
Equation \ref{eq24} requires 
\begin{equation}
\Omega =\exp \left[ \Phi _N\left( x^\mu \right) \right] \text{ ,}
\label{eq34}
\end{equation}
where $m\left( x^\mu \right) $ is the mass of a fundamental particle in the
JF and $\widetilde{m}_0$ its invariant mass in the EF. The conformal
equivalence with canonical GR is a consequence of the coupling constant 
\[
\omega =-\frac 32\text{ ,} 
\]
and then defining the EF by $G$ $=\,G_N$ a constant. This value for $\omega$  
may simply be set empirically, but it can be shown to be required from first principles 
(Barber, 2003).

\subsection{The SCC Field Equations}

The result of these three requirements gave the following fundamental,
manifestly covariant, field equations:

The scalar field equation 
\begin{equation}
\Box \phi =4\pi T_M^{\;}\text{ ,}  \label{eq143}
\end{equation}
the gravitational field equation

\begin{eqnarray}
R_{\mu \nu }-\frac 12g_{\mu \nu }R &=&\frac{8\pi }\phi T_{M\mu \nu }-\frac
3{2\phi ^2}\left( \nabla _\mu \phi \nabla _\nu \phi -\frac 12g_{\mu \nu
}g^{\alpha \beta }\nabla _\alpha \phi \nabla _\beta \phi \right)
\label{eq143a} \\
&&\ \ +\frac 1\phi \left( \nabla _\mu \nabla _\nu \phi -g_{\mu \nu }\Box
\phi \right) \text{ ,}  \nonumber
\end{eqnarray}
and the creation equation, which replaces the conservation equation
(Equation \ref{eq1})

\begin{equation}
\nabla _\mu T_{M\;\nu }^{\;\mu }=\frac 1{8\pi }\frac 1\phi \nabla _\nu \phi
\Box \phi \text{ .}  \label{eq144}
\end{equation}

\subsection{The Static, Spherically Symmetric Solution}

The Robertson parameters are 
\begin{equation}
\alpha _r=1\qquad \beta _r=1\qquad \gamma _r=\frac 13\text{ ,}  \label{eq145}
\end{equation}
and therefore the standard form of the Schwarzschild metric is 
\begin{eqnarray}
d\tau ^2 &=&\left( 1-\frac{3G_NM}r+..\right) dt^2-\left( 1+\frac{G_NM}%
r+..\right) dr^2  \label{eq146} \\
&&\ \ \ \ \ \ \ \ -r^2d\theta ^2-r^2\sin ^2\theta d\varphi ^2\text{ .} 
\nonumber
\end{eqnarray}
The formula for $\phi $ is 
\begin{equation}
\phi =G_N^{-1}\exp (-\Phi _N)  \label{eq108}
\end{equation}
and that for $m$ is, (Equation \ref{eq24}), 
\[
m_p\left( x_\mu \right) =m_0\exp (\Phi _N)\text{.} 
\]

The effect of breaking the equivalence principle in accordance with the PMI
is that there is an extra scalar field force, which acts on particles but
not photons, that behaves as Newtonian gravitation except in the opposite
direction. There are therefore two gravitational constants, one felt by
photons, $G_m$ , is that describing the curvature of space-time and the
other $G_N$ is that felt by particles and is the normal Newtonian
gravitational constant measured in Cavendish type experiments.

A detailed calculation yields (Barber, 2002a)
\begin{equation}
G_N=\frac 23G_m\text{ ,}  \label{eq149}
\end{equation}
so the acceleration of a massive body caused by the curvature of space-time
is $\frac 32$ the Newtonian gravitational acceleration actually experienced.
However this is compensated by an opposite acceleration of $\frac 12$
Newtonian gravity due to the scalar field.

The composite curvature and scalar field accelerations of a freely falling
particle measured in the rest frame of the Centre of Mass (CoM) frame of
reference is 
\begin{equation}
\frac{d^2r}{dt^2}=-\left\{ 1-\frac{G_NM}r+...\right\} \frac{G_NM}{r^2}\text{
.}  \label{eq150}
\end{equation}
and the forces acting on a freely falling particle as measured in that rest
frame are 
\begin{equation}
m_0\frac{d^2r}{dt^2}=-m(r)\frac{G_NM}{r^2}\text{ ,}  \label{eq151}
\end{equation}
$m_0$ can be thought of as ''inertial-mass'', which measures inertia and $%
m(r)$ as ''gravitational mass'', which interacts with the gravitational
field with $\stackunder{r\rightarrow \infty }{Lim}\,m(r)=m_{0\text{ }}$.

\section{Experimental Consequences of the Theory}

\subsection{The Gravitational Red Shift of Light}

In SCC the principle of the local conservation of energy was applied to the
gravitational red shift of light. The analysis depended on the assumption
that if no work is done on, or by, a projectile while in free fall then its
energy $E$ , $P^0$ , is conserved when measured in a specific frame of
reference, that of the CoM of the system.\textit{\ }

When a photon is emitted by one atom at altitude $x_2$ and absorbed by
another at an altitude $x_1$ , the standard time dilation relationship is 
\begin{equation}
\frac{\nu \left( x_2\right) }{\nu \left( x_1\right) }=\left[ \frac{%
-g_{00}\left( x_2\right) }{-g_{00}\left( x_1\right) }\right] ^{\frac 12}%
\text{ .}  \label{eq39}
\end{equation}
If $x_2=r$ and $x_1=\infty $ , where $g_{00}\left( x_1\right) =-1$, and
writing $\stackunder{r\rightarrow \infty }{Lim}\,\nu \left( r\right) $ as $%
\nu _0$, the standard (GR) gravitational red shift relationship is derived 
\begin{equation}
\nu \left( r\right) =\nu _0\left[ -g_{00}\left( r\right) \right] ^{\frac 12}%
\text{ ,}  \label{eq40}
\end{equation}
where the observer is at infinite altitude observing a photon emitted at
altitude $r$. The next step was to consider the rest mass, $m\left( r\right) 
$, of a projectile launched up to an altitude $r$ while locally conserving
energy, the rest mass was evaluated in the co-moving CoM frame as 
\begin{equation}
m_c\left( r\right) =m_0\exp \left[ \Phi _N\left( r\right) \right] \left[
-g_{00}\left( r\right) \right] ^{\frac 12}\text{ ,}  \label{eq63}
\end{equation}
where the observer is at infinite altitude 'looking down' to a similar
particle at an altitude $r$. From this expression it was obvious that with
the assumption of the conservation of energy, $P^0$ , in the CoM frame,
gravitational time dilation, the factor $\left[ -g_{00}\left( r\right)
\right] ^{\frac 12}$, applies to massive particles as well as to photons.

As physical experiments measuring the frequency of a photon compare its
energy with the mass of the atom it interacts with, it is necessary to
compare the masses (defined by Equation \ref{eq63}) of two atoms at
altitude, $r$ and $\infty $, with the energy (given by Equation \ref{eq40})
of a ''reference'' photon transmitted between them. This yielded the
physical rest mass $m_p\left( r\right) $ as a function of altitude 
\begin{equation}
\frac{m_p\left( r\right) }{\nu \left( r\right) }=\frac{m_0}{\nu _0}\exp
\left[ \Phi _N\left( r\right) \right] \text{ .}  \label{eq64}
\end{equation}

Equation \ref{eq64} is a result relating observable quantities, but how is
it to be interpreted? In other words how are mass and frequency to be
measured in any particular frame? In the GR EF (and BD JF) the physical rest
mass of the atom is defined to be constant, hence prescribing ($\widetilde{x}%
^\mu $), with $m_p\left( \widetilde{r}\right) =m_0$. In this case Equation 
\ref{eq64} becomes 
\begin{equation}
\nu \left( \widetilde{r}\right) =\nu _0\left( 1-\widetilde{\Phi }_N\left( 
\widetilde{r}\right) +...\right) \text{ .}  \label{eq65}
\end{equation}

However in the SCC JF(E) rest mass is given by the expression Equation \ref
{eq24}, consequently a comparison of Equation \ref{eq64} with the equation
for rest mass in this frame yields 
\begin{equation}
\nu \left( r\right) =\nu _0\text{ .}  \label{eq68}
\end{equation}
In the JF(E) the energy of a photons is conserved, even when transversing
curved space-time. Gravitational red shift is interpreted as a gain of
potential energy, and hence mass, of the apparatus, rather than a loss of
(potential) energy of the photon.

Using either frame the gravitational red shift prediction in SCC is in
agreement with GR and all observations to date .

\subsection{The Observational Tests of SCC}

The three original ''classical'' tests of GR suggested by Einstein; the
deflection of light by the sun, the gravitational red shift of light and the
precession of the perihelia of the orbit of Mercury, together with the time
delay of radar echoes passing the sun, the precession of a gyroscope in
earth orbit and the ''test-bed'' of GR, the binary pulsar PSR 1913 + 16 will
now be examined in the SCC JF. In order to demonstrate the conformal
equivalence of SCC JF with GR the parameter $\lambda $ is initially left
undetermined.

Now with a general $\lambda $ the relationship between $G_m$ and $G_N$ was
found to be 
\begin{equation}
G_m=\frac{\left( 2+\lambda \right) }2G_N\text{ ,}  \label{eq127}
\end{equation}
and the Robertson parameter $\gamma _r$ is given by 
\begin{equation}
\gamma _r=\frac{\left( 2-\lambda \right) }{\left( 2+\lambda \right) }
\label{eq137}
\end{equation}
In several classical tests, using the Robertson parameters, a factor $\Gamma 
$ appears where: 
\begin{equation}
\Gamma =\left( \frac{1+\gamma _r}2\right) G_m\text{ }  \label{eq177}
\end{equation}
and substituting for $\gamma _r$ and $G_m$ it is found that whatever the
value of $\lambda $ 
\begin{equation}
\Gamma =G_N\text{ .}  \label{eq177a}
\end{equation}

\subsubsection{The Deflection of Light.}

The Robertson parameter expression for the deflection of light by a massive
body is 
\begin{equation}
\theta =\frac{4G_mM}R\left( \frac{1+\gamma _r}2\right) =\frac{4G_NM}R\text{ ,%
}  \label{eq178}
\end{equation}
so for the sun $\theta =1.75"$ in exact agreement with GR and observation.

The deflection, Equation \ref{eq178} may be divided into two components 
\[
\theta =\frac{4G_mM}R\left( \frac 12\right) +\frac{4G_mM}R\left( \frac{%
\gamma _r}2\right) \text{ ,} 
\]
the deflection consists of a 'gravitational attraction' of the photon, the
effect of the equivalence principle, and an extra deflection caused by
curvature. In canonical GR, where $G_m=G_N$ and $\gamma =1$, the first
component is equal to the second and both equal $\frac{2G_NM}R$ . On the
other hand in SCC, $G_m=\frac 32G_N$ , and so the first component is equal
to $\frac{3G_NM}R$ , however as curvature is reduced by a factor $\gamma
=\frac 13$ , the second component is only $\frac{G_NM}R$ , thus resulting in
a total deflection of $\frac{4G_NM}R$ , equal to that of GR.

\subsubsection{Radar Echo Delay}

The delay in the timing of radar echoes passing the sun and reflected off
(say) Mercury at superior conjunction provides a further test for the $%
\gamma $ Robertson parameter. The expression for the delay is given by
Misner et al. (Misner, Thorne \& Wheeler, 1973) as 
\begin{equation}
\frac{d\triangle \tau }{d\tau }-(\text{Constant Newtonian part})=-4\left(
1+\gamma _r^{GR}\right) \frac{GM}b\frac{db}{d\tau }  \label{eq179}
\end{equation}
(where$\ $ $b$ is the distance of the ray from the earth-sun axis), and
experiments have shown $\gamma _r^{GR}=1$ to a high degree of accuracy. In
SCC 
\begin{equation}
\frac{d\triangle \tau }{d\tau }-(\text{Constant Newtonian part})=-8\frac{%
\left( 1+\gamma _r\right) }2\frac{G_mM}b\frac{db}{d\tau }=-8\frac{\Gamma M}b%
\frac{db}{d\tau }  \label{eq180}
\end{equation}

therefore, as $\Gamma =G_N$, SCC again predicts the same result as GR.

\subsubsection{The Precession of the Perihelia}

The precession of perihelia of an orbiting body, primarily the planet
Mercury, does not depend on the parameter $\Gamma $. It is given in terms of
the Robertson parameters as 
\begin{equation}
\triangle \theta =\left( \frac{6\pi G_mM}L\right) \left( \frac{2-\beta
_r+2\gamma _r}3\right) \qquad \text{radians/rev.}  \label{eq183}
\end{equation}
where $L$ is the semilatus rectum. In SCC, as in BD $\beta _r=1$ and $\gamma
_r=\frac{\left( 2-\lambda \right) }{\left( 2+\lambda \right) }$ this yields 
\begin{equation}
\triangle \theta =\left( 1-\frac \lambda 6\right) \left( \frac{6\pi G_NM}%
L\right) \qquad \text{radians/rev.}  \label{eq184}
\end{equation}

However we also have to allow for the effect of the action of the scalar
field which modifies Newtonian gravitation according to Equation \ref{eq150}%
, using units with $c\neq 1$ this becomes, 
\begin{equation}
\frac{d^{2}r}{dt^{2}}=-\left[ 1-\lambda \frac{G_{N}M}{rc^{2}}\right] \frac{%
G_{N}M}{r^{2}}\text{ .}  \label{eq185}
\end{equation}

This can be considered as the acceleration produced by a Newtonian potential
with a dipole-like perturbing potential 
\begin{equation}
\Phi =-\frac{G_NM}r+\frac 12\lambda \left( \frac{G_NM}{rc}\right) ^2
\label{eq186}
\end{equation}
This non-Newtonian perturbation produces an extra precession of a factor 
\[
\frac \lambda 6 
\]
of the full GR perihelion advance. Combining this extra perihelion advance
with that caused by the curvature of space-time in Equation \ref{eq184}
results in a SCC prediction of a PNA perihelion advance equal to 
\begin{equation}
\triangle \theta =\frac{6\pi G_NM}L\qquad \text{radians/rev.}  \label{eq190}
\end{equation}
So there is in exact agreement with the canonical GR value. Note this
additional precession is that of the 'semi-relativistic' adaptation of the
mass that includes potential energy in Newtonian orbital dynamics.

\subsubsection{The Binary Pulsar PSR 1913 + 16}

A neutron star, composed of relativistic matter with an equation of state of 
\[
p_n=\frac 13\rho _n
\]
will be de-coupled from the scalar field. Without further analysis it seems
likely that any predictions of Binary pulsar loss of orbital energy due to
gravitational radiation will be the same as GR. However in the formation of
a collapsed star the gravitational field would appear to increase by a
factor of $1.5$ as the gravitating mass became degenerate and de-coupled
from the scalar field.

\subsubsection{The Precession of a Gyroscope}

The effect of the curvature of space-time on the precession of a gyroscope
in earth orbit is similarly compensated for by the scalar field. The
component $\stackrel{3}{g}_{i0}$ was calculated to be ([Barber,2002a),
(Weinberg,1972): 
\[
\nabla ^2\stackrel{3}{g}_{i0}=16\pi G_m\left( \frac{2\varpi +3}{2\varpi +4}%
\right) \stackrel{1}{T}^{i0}+\left( \frac 2{\varpi +2}\right) \frac{d^2\Phi
_m}{dx^idt}\text{ ,} 
\]
and therefore for a static system 
\begin{equation}
\stackrel{3}{g}_{i0}=-4G_m\left( \frac 2{2+\lambda }\right) \int \frac{%
\stackrel{1}{T}^{i0}(x^{\prime },t)}{\left| x-x^{\prime }\right| }d^3x\text{
.}  \label{eq181}
\end{equation}
Now 
\begin{equation}
G_m\left( \frac 2{2+\lambda }\right) =\Gamma  \label{eq182}
\end{equation}
As $\Gamma =G_N$ , the effects of the rotation of the earth, or any central
spherical mass, on the precession of spins and perihelia are the same in
this theory as in canonical GR. (GR may be obtained by letting $\lambda
\rightarrow 0$ in Equation \ref{eq181}.) Hence in the Gravity Probe B
Lense-Thirring experiment SCC predicts the identical result as GR.

\section{The Definitive Experiments}

\subsection{Do Photons fall at the same rate as Particles?}

The identical predictions in the One-Body Problem in GR and SCC raise the
question ''Is SCC just GR rewritten in some obscure coordinate system, the
JF(E) rather than the EF?'' That this is not so and SCC is indeed a separate
theory from GR may be seen when the behaviour of light is compared with that
of matter in free fall. Although the prediction of the deflection of light
by massive bodies is equal in both theories, in SCC a photon in free fall
descends at $\frac 32$ the acceleration of matter. i.e. in free fall a beam
of light travelling a distance $l$ is deflected downwards, relative to
physical apparatus, by an amount 
\begin{equation}
\delta =\frac 14g\left( \frac lc\right) ^2\text{ .}  \label{eq191}
\end{equation}

As a possible experiment I suggest launching into earth orbit an annulus,
two meters in diameter, supporting 1,000 carefully aligned small mirrors. A
laser beam is then split, one half reflected, say 1,000 times, to be
returned and recombined with the other half beam, reflected just once, to
form an interferometer at source. If the experiment is in earth orbit and
the annulus orientated on a fixed star, initially orthogonal to the orbital
plane then the gravitational or acceleration stresses on the frame, would
vanish whereas they would predominate on earth. In orbit SCC predicts a 2
Angstrom interference pattern shift with a periodicity equal to the orbital
period whereas GR predicts a null result.

\subsection{ Is there a Cut-Off to the Casimir Force?}

In the JF gravitational acceleration, Equation \ref{eq151}, can be expressed
as 
\begin{equation}
\frac{d^2r}{dt^2}=-\exp \left( \Phi _N\right) \frac{G_NM}{r^2}\text{ , }
\label{eq152}
\end{equation}
and hence it followed that 
\begin{equation}
\nabla \left[ \exp \left( -\Phi _N\right) \right] =-\frac{G_NM}{r^2}\text{ }
\label{eq153}
\end{equation}
so, 
\begin{equation}
\Phi _N=-\ln \left( 1+\frac{G_NM}r\right) \text{ .}  \label{eq154}
\end{equation}

On the other hand in the EF $\widetilde{m}(\widetilde{r})=m_0$ , therefore
Equation \ref{eq151} reduces to the normal Newtonian/GR expression 
\begin{equation}
\frac{d^2\widetilde{r}}{d\widetilde{t}^2}=-\frac{G_N\widetilde{M}}{%
\widetilde{r}^2}\text{ .}  \label{eq156}
\end{equation}
This is derived, of course, from the usual EF Newtonian potential 
\begin{equation}
\widetilde{\Phi }_N=-\frac{G_N\widetilde{M}}{\widetilde{r}}\text{ .}
\label{eq157}
\end{equation}
The difference in the Newtonian potentials for the two frames is the
consequence of the SCC EF having a classical Lagrangian and the JF(E) having
a non-classical Lagrangian \textit{in vacuo}. This difference between the
classical EF and non-classical JF(E) manifests itself in the vacuum solution
to the field equations. In the JF(E) the Newtonian potential solution,
obtained from the principle of the conservation of energy, requires an
additional traceless potential of 
\begin{equation}
\nabla \Phi _N=-\left[ \frac{G_NM}r\right] ^2  \label{eq167}
\end{equation}
to that of the vacuum solution derived from the Principle of Mutual
Interaction. This is the Newtonian potential of a small ''quantum ether''
vacuum density $\rho _{qv}$ where 
\begin{equation}
\rho _{qv}=-\frac 1{2\pi }\frac{G_NM}r\frac M{r^3}\text{ .}  \label{eq168}
\end{equation}
Furthermore, introducing $\rho _{av}$ as the average matter density inside
the sphere, radius $r$, centered on the gravitating mass $M$, this can be
written as 
\begin{equation}
\rho _{qv}=-\frac 23\frac{G_NM}r\rho _{av}\text{ .}  \label{eq169}
\end{equation}
The negative sign is consistent with standard analysis of the Casimir effect
in which the ''quantum ether'' between the Casimir conductors has a negative
energy density. So in a laboratory near the earth 
\begin{equation}
\rho _{qv\oplus }\simeq -2.4\times 10^{-9}\text{ gm.cm}^{-3}\text{ .}
\label{eq170}
\end{equation}
This density is proportional to $r^{-4}$ and limits the maximum Casimir
force that might be detected. This limit may be detectable at a sufficient
distance from gravitational masses. Thus the theory implies that in flat
space-time, in the absence of gravitational fields, the Casimir force would
not be detectable at all! The theory does suggest that an experiment
launched away from the sun, which compared the Casimir force against
separation, would detect the force rounding off as the limit to the Casimir
effect was reached. This limit may be detected at around 5 A.U. with current
experimental sensitivity.

\subsection{Geodetic Precession}

The Gravity Probe B experiment, successfully completed in September 2005 
and now in a prolonged data analysis phase, has compared the spin directions 
of an array of four redundant gyroscopes. It is testing the Lense-Thirring or 
frame-dragging effect, in which the SCC prediction is equal to that of GR 
as above in Equation \ref{eq181}. This is a value of 0.042 arc/yr about a 
direction parallel to the direction of the Earth's rotation axis. 
(Keiser,G.M., et al, 2002). The interesting aspect from the SCC point of view 
is that the experiment is also measuring the geodetic effect. This effect is 
described by the expression (Will, 2002) 
\begin{equation}
\frac 12\left( 2\gamma +1\right) \frac{GM_{\oplus }^{}}{R^3}\mathbf{v}%
_s\times \mathbf{X}  \label{eq171}
\end{equation}
which in GR, where $\gamma =1$ and $G=G_N$ , predicts a precession for the
Gravity B Probe gyroscope of 6.6 arc sec/yr about a direction perpendicular
to the plane of the orbit, this is given by, 
\begin{equation}
\Omega = \mathbf{v}\times[-\frac{1}{2}\mathbf{a}+(\gamma+\frac{1}{2})\mathbf{\nabla}U] \text,\label{eq172}
\end{equation}
where $\mathbf{a}$ is the acceleration from the geodesic and $U = - \Phi$ is the 'Newtonian' 
potential of the metric being considered.

In SCC $\gamma =\frac 13$ and $G=G_m=\frac 32G_N$, i.e. $\mathbf{\nabla}U_m = \frac32\mathbf{\nabla}U_N$, 
so the theory predicts a geodetic precession of 5/6 of the GR geodetic precession or just 
5.5 arc sec/yr. However, a further correction has to be made to the SCC prediction. The 
satellite in drag-free mode in orbit is not travelling along a geodesic of the SCC metric but 
perturbed from it by the scalar field force. The inertial acceleration produced by this force 
is given by, $\mathbf{a} = \frac13\mathbf{\nabla}U$. The Thomas precession correction is thus 
$\mathbf{v}\times1/6\mathbf{\nabla}U$, i.e. -1/6 the GR geodetic precession. Therefore the 
SCC prediction of the total precession about a direction perpendicular to the plane of the orbit 
is 2/3 the GR value. That is SCC predicts a N-S precession of 4.4096 arcsec/yr.

\section{Conclusions}

\subsection{A Summary}

Two key aspects of the theory are
Firstly it is not a classically metric theory, it is a 'semi-metric' theory in which:  

 		-  photons follow geodesics; but particles do not, the Principle of Equivalence is replaced by the Principle
of Mutual Interaction.

Secondly, there is a conformal equivalence \textit{in vacuo} between the Jordan frame and GR in its 
Einstein frame,

		- the JF(E) describes curvature and conserves mass-energy

		and in the EF, which is canonical General Relativity \textit{in vacuo}, 
four-energy-momentum is conserved. 

Consequently test bodies falling freely \textit{in vacuo} experience a scalar field force that 
exactly compensates for the effect of the scalar field on curvature. Particles follow 
GR geodesics \textit{in vacuo}. \textit{In vacuo} there is a degeneracy between GR and SCC
that is only resolved in the definitive experiments described above. In the cosmological solution
where there is a homogeneous density the solution does differ from the standard GR solution
and yields a concordant cosmological model that does not require the unverified physics of inflation, 
exotic Dark Matter or Dark Energy.
 
As calculated in the earlier paper (Barber, 2002a), the gravitational constant $G_m$, that 
determines the coupling  of matter to curvature, is greater by a factor $\frac 32$ from that 
measured as Newtonian $G_N$ in Cavendish type experiments. As shown above this increase 
compensates for the reduced value of $\gamma =\frac 13$. Thus, using Newtonian $G_N$,  
interpretations of the data in the deflection of light, frame dragging, and radar echo delay 
experiments would determine a $\gamma = 1$.

The geodetic measurement, which considers the Earth - Moon system as a gyroscope, would also 
appear to be similarly compensated. This is because, as it is an extended, gravitationally bound 
system \textit{in vacuo}, the problem can be considered in the Einstein frame of the theory, 
which is canonical GR. The Earth and Moon follow their GR geodesic trajectories through 
space-time.

As a result of this conformal equivalence between the two theories it has not been possible to 
distinguish SCC from GR in all previous solar system experiments, there is a degeneracy in these 
tests between the two theories. There are the two possible further experiments as suggested above, 
which would distinguish between them. But also, there is the third experiment being evaluated at present, 
the Gravity Probe B satellite, which is able to differentiate between the two theories and 
resolve the degeneracy.

As the gyroscopes in the Gravity Probe B experiment are solid and their interiors not 
'\textit{in vacuo}' the experiment cannot be conformally transformed into a canonical 
GR Einstein frame. It is a 'point' measurement of curvature and as such it is the first 
non-null experiment to distinguish between GR and SCC. As the results of this experiment are 
about to be published in 2006/7, it will imminently provide the first occasion to test 
SCC against GR and therefore this experiment presents an important opportunity to 
falsify the theory. 

\subsection{The prediction}

In the Gravity Probe B satellite experiment SCC and GR predict gyroscope precessions, 
about a direction perpendicular to the plane of the orbit, of 4.4096 arcsec/yr
and 6.6144 arc sec/yr respectively.

\subsection{Acknowledgments}

I am thankful for an earlier discussion with Professor Bernard Carr, Queen Mary and
Westfield College, University of London. Any errors and misconceptions remaining are of 
course entirely my own.

\end{document}